%% file: main.tex
\documentclass[aps ,prb, reprint, floatfix,nobibnotes,amsmath, amssymb,superscriptaddress, flushbottom]{revtex4-2}

\usepackage{lipsum}
\usepackage{graphicx}
\usepackage[unicode=true, colorlinks=true, citecolor={blue!80!black}, urlcolor={blue!50!black}, linkcolor = {blue!80!black}]{hyperref}
\usepackage[normalem]{ulem}
\usepackage{placeins} 
\usepackage{afterpage}
\usepackage{titlesec}
\usepackage[svgnames]{xcolor}

\newcounter{appendixpar}
\renewcommand{\theappendixpar}{Appendix~\Alph{appendixpar}}

\newcommand{\appendixpar}[1]{%
  \refstepcounter{appendixpar}%
  \textit{\theappendixpar}: \textit{#1}.---%
}

\newcommand{\cpar}[1]{%
    \textit{#1}.---%
}

\newif\ifSMincluded
\SMincludedfalse

\SMincludedtrue

\newcommand{\SMcitekey}{SupplementalMaterial}  

\newcommand{\SMref}[2]{%
  \ifSMincluded
    \hyperref[#1]{#2}%
  \else
    #2~\cite{\SMcitekey}%
  \fi
}

\begin{document}

\preprint{APS/123-QED}

\title{Probing synthetic Caroli-de Gennes-Matricon states through critical current in full-shell nanowire Josephson junctions}

\author{Carlos Payá}
\thanks{These authors have contributed equally to this work.}
\affiliation{Instituto de Ciencia de Materiales de Madrid (ICMM), CSIC, 28049 Madrid, Spain}
\author{\'{A}ngel Ibabe}
\thanks{These authors have contributed equally to this work.}
\affiliation{Departamento de F\'{i}sica de la Materia Condensada, Universidad Aut\'{o}noma de Madrid, Madrid, Spain}
\affiliation{Condensed Matter Physics Center (IFIMAC), Universidad Aut\'{o}noma de Madrid, Madrid, Spain}
\author{Mario Gómez}
\thanks{These authors have contributed equally to this work.}
\affiliation{Departamento de F\'{i}sica de la Materia Condensada, Universidad Aut\'{o}noma de Madrid, Madrid, Spain}
\affiliation{Condensed Matter Physics Center (IFIMAC), Universidad Aut\'{o}noma de Madrid, Madrid, Spain}
\author{Thomas Kanne}
\affiliation{Center for Quantum Devices, Niels Bohr Institute, University of Copenhagen, Copenhagen, Denmark}
\author{Jesper Nyg\r{a}rd}
\affiliation{Center for Quantum Devices, Niels Bohr Institute, University of Copenhagen, Copenhagen, Denmark}
\author{Ramón Aguado}
\affiliation{Instituto de Ciencia de Materiales de Madrid (ICMM), CSIC, 28049 Madrid, Spain}
\affiliation{Quantum Advanced Research Center (QuARC), CSIC, Madrid, Spain}
\affiliation{Laboratorio de Transporte Cu\'{a}ntico, Unidad Asociada
UAM/ICMM-CSIC, Madrid, Spain}
\author{Pablo San-Jose}
\affiliation{Instituto de Ciencia de Materiales de Madrid (ICMM), CSIC, 28049 Madrid, Spain}
\affiliation{Laboratorio de Transporte Cu\'{a}ntico, Unidad Asociada
UAM/ICMM-CSIC, Madrid, Spain}
\author{Eduardo J. H. Lee}
\email{eduardo.lee@uam.es}
\affiliation{Departamento de F\'{i}sica de la Materia Condensada, Universidad Aut\'{o}noma de Madrid, Madrid, Spain}
\affiliation{Condensed Matter Physics Center (IFIMAC), Universidad Aut\'{o}noma de Madrid, Madrid, Spain}
\affiliation{Laboratorio de Transporte Cu\'{a}ntico, Unidad Asociada
UAM/ICMM-CSIC, Madrid, Spain}
\affiliation{Instituto Nicolás Cabrera (INC), Universidad Aut\'{o}noma de Madrid, 28049 Madrid, Spain}
\author{Elsa Prada}
\email{elsa.prada@csic.es}
\affiliation{Instituto de Ciencia de Materiales de Madrid (ICMM), CSIC, 28049 Madrid, Spain}
\affiliation{Laboratorio de Transporte Cu\'{a}ntico, Unidad Asociada
UAM/ICMM-CSIC, Madrid, Spain}

\date{\today}
      
\begin{abstract}
Full-shell hybrid nanowires consisting of a semiconductor core fully enveloped by a superconducting shell have emerged as a platform to study Caroli-de Gennes-Matricon (CdGM) analogs. These subgap states can be considered a synthetic version of CdGM states in Abrikosov vortices. Unlike conventional CdGM states, these analogs exhibit a level spacing comparable to the superconducting gap, making them readily observable via tunneling spectroscopy techniques. The spectral density of CdGM analogs follows a characteristic skewed pattern as a function of applied axial magnetic field, an effect that is superimposed on the Little-Parks oscillations of the shell's gap induced by fluxoid quantization. Here, we provide experimental evidence for CdGM analogs through a distinctive skewness fingerprint in the critical current and zero-bias resistance of overdamped Josephson junctions based on full-shell nanowires. 
\end{abstract}

\maketitle

Caroli-de Gennes-Matricon (CdGM) states are electronic states bound to the core of Abrikosov vortices in type-II superconductors (SCs) \cite{Caroli:PL64,Tinkham:96}. Their energy level spacing is typically so small compared to the SC gap that they are notoriously difficult to resolve \cite{Berthod:PRL17, Chen:NC18, Chen:PRL20}. A synthetic analog of the Abrikosov vortex has emerged in the form of a full-shell hybrid nanowire. This system consists of a semiconductor (SM) core fully enveloped by a thin SC shell and subjected to an axial magnetic field. Inside these wires, subgap bound states also appear that share some similarities to CdGM states but also important differences \cite{San-Jose:PRB23,Deng:PRL25}. They are thus dubbed CdGM analogs. 

In recent years, full-shell hybrid nanowires have been extensively investigated, particularly regarding the phenomenology of the shell gap \cite{Razmadze:PRL20, Sabonis:PRL20, Kringhoj:PRL21, Ibabe:NC23, Razmadze:PRB24a, Ibabe:NL24}, subgap states \cite{Kopasov:PRB20, Kopasov:PSS20, Escribano:PRB22, San-Jose:PRB23, Giavaras:PRB24}, and their potential as topological SCs \cite{Vaitiekenas:S20,Penaranda:PRR20, Valentini:S21, Valentini:N22, Paya:PRB24, Paya:PRB24a}. In the presence of magnetic flux, the doubly-connected geometry of the shell enforces an integer phase winding $n$ of the SC order parameter, also known as the fluxoid number \cite{London:50, Tinkham:96}. This quantization gives rise to the Little-Parks (LP) effect \cite{Little:PRL62, Parks:PR64}, in which the shell gap oscillates with the applied flux, forming a lobe structure.

Inside this gap, the spectral density of full-shell nanowires is dominated by the CdGM analogs. Their behavior as a function of flux has an overall LP modulation, but within each $n\neq 0$ lobe, they display a characteristic \textit{skewness} with flux reflecting the interplay of the orbital coupling and the radial confinement of the core states \cite{San-Jose:PRB23}. Importantly, the level spacing of CdGM analogs turns out to be comparable to the SC gap \cite{San-Jose:PRB23} and thus easier to detect experimentally. The presence of CdGM analogs in full-shell nanowires was demonstrated via tunneling spectroscopy in Ref.~\cite{Deng:PRL25}. Subsequent theoretical works showed that a Josephson junction between two full-shell nanowires inherits this asymmetry, resulting in a skewed critical current directly tied to the underlying CdGM analogs \cite{Paya:PRB25, Paya:PRB25a}. However, while full-shell Josephson junctions have been explored experimentally \cite{Razmadze:PRL20, Sabonis:PRL20, Ibabe:NC23, Razmadze:PRB23, Ibabe:NL24}, a clear connection between the Josephson current and the presence of CdGM analogs has not yet been established. 

\begin{figure}
\includegraphics{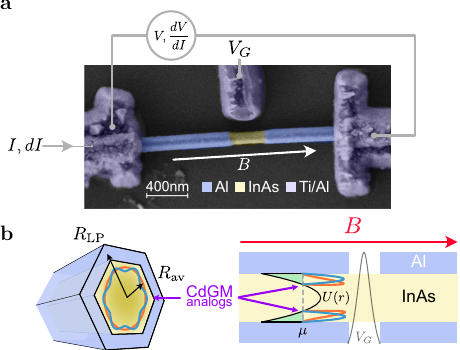}
\caption{\label{fig:sketch} 
{\textbf{Full-shell nanowire Josephson junction.}} (a) False-color electron micrograph of a representative device. The junction is defined by selectively etching the Al shell from the central region of an Al/InAs nanowire, and contacting the ends with Ti/Al leads. Transport is characterized by applying a DC current bias $I$, mixed with a small AC excitation, while simultaneously measuring the voltage drop $V$, and the differential resistance $dV/dI$, with an external magnetic field $B$ applied approximately aligned to the nanowire axis. A local gate voltage $V_G$ at the junction is used to tune its transparency. (b) Schematic of the theoretical model, showing a full-shell nanowire cross-section and a Josephson junction. For a given chemical potential $\mu$, the radial electrostatic potential $U(r)$ localizes Caroli–de Gennes–Matricon (CdGM) analogs (orange/dark blue) at an average radius $R_{\rm{av}}$ smaller than the shell average radius $R_{\rm{LP}}$.
}
\end{figure}

In this work, we measure the differential resistance of an overdamped full-shell Josephson junction as a function of applied bias current $I$ and an axial magnetic field $B$, see Fig. \ref{fig:sketch}(a). In an overdamped junction, the differential resistance is predicted to peak at an applied current $I\approx I_c$, where $I_c$ is the intrinsic critical current of the junction \cite{Tinkham:96}. We find that this $I_c$ is modulated by the LP effect but is also distinctly skewed within each $n\neq 0$ lobe. Additionally, we present the zero-bias resistance normalized to the  normal resistance ($R_0/R_N$).  This ratio is similarly skewed and yields an independent measurement of $I_c$ via the Ambegaokar-Halperin relation for overdamped junctions in the thermal regime \cite{Ambegaokar:PRL69, Gross:16}. We compare both experimental features to microscopic numerical simulations of $I_c$ using the theoretical model schematically depicted in Fig. \ref{fig:sketch}(b). We find good agreement between them, demonstrating that the skewness is a direct consequence of CdGM analogs residing within the nanowire.



We fabricated a device consisting of a Josephson junction defined in an InAs nanowire fully coated with an Al shell, illustrated in Fig. \ref{fig:sketch}(a). The junction transparency is controlled by a local gate $V_G$, with $B$ applied nearly parallel to the wire. Two superconducting Ti/Al electrodes were deposited on top of the Al shell to perform current-biased transport measurements using
standard lock-in techniques. A DC current bias $I$, mixed with a small low-frequency AC excitation $dI$, was applied while simultaneously measuring the voltage drop $V$ and the differential resistance $dV/dI$. For further fabrication and measurement details, see \ref{ap:fabrication} and \ref{ap:measurements}, respectively, and Supplemental Material \SMref{SM:device}{S-I}.  

\begin{figure} 
\centering
\includegraphics[width=\columnwidth]{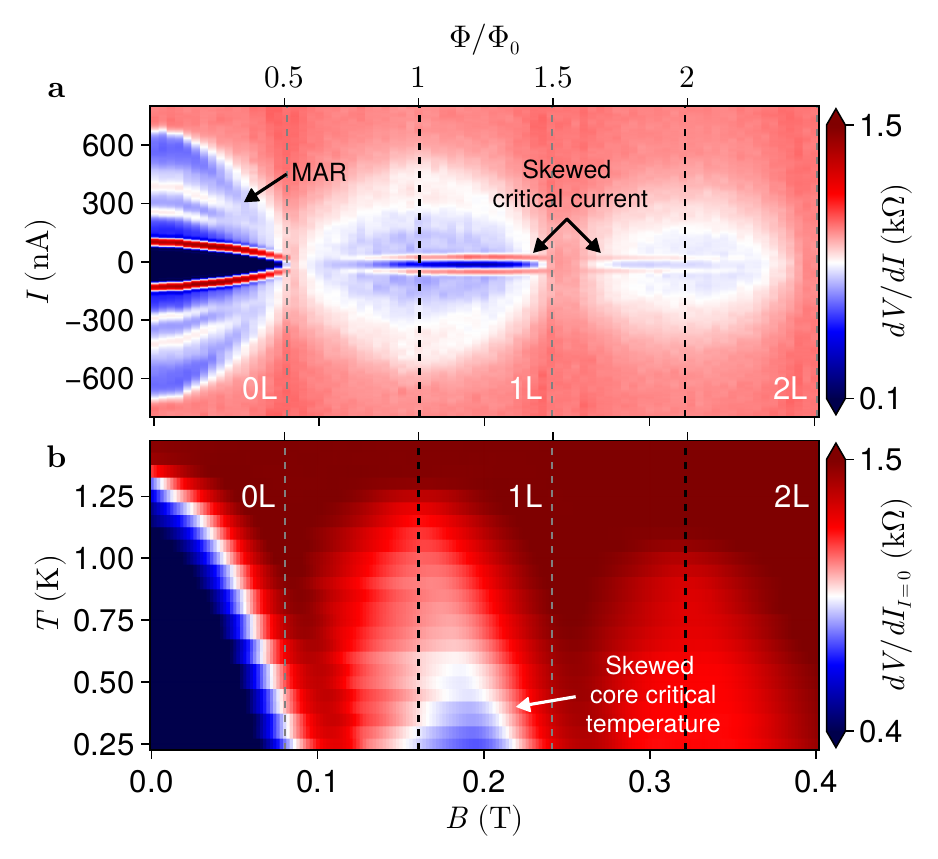} 
\caption {\label{fig:exp}
{\textbf{Skewness of the critical current and critical temperature.}} (a) Differential resistance $dV/dI$ measured as a function of axial magnetic field $B$ and applied current $I$. The outer-most features oscillate with the flux $\Phi$ threading the shell, forming Little-Parks (LP) lobes labeled by $n$L, where $n$ is the fluxoid number. $\Phi_0$ is the superconducting flux quantum. The critical current $I_c$ at low current bias presents a clear skewness towards high (low) magnetic fields within 1L (2L). Multiple Andreev reflection (MAR) peaks are visible in 0L. (b) Zero-bias $dV/dI$ measurement as a function of $B$ and cryostat nominal temperature $T$. The high-temperature features oscillate with flux owing to the LP effect of the shell, while the critical temperature of the hybridized core is skewed in 1L. Panel (a) is taken at $T=250$mK.
}
\end{figure}

The main transport features are shown in Fig. \ref{fig:exp}(a), where $dV/dI$ is measured as a function of $B$ and $I$ at a fixed gate voltage $V_G = 80$~V. 
The $I-V$ characteristics are non-hysteretic, i.e., single valued and reversible, indicating that the junction is in the overdamped regime \cite{Tinkham:96}.
The outer-most features at high bias (in white color) oscillate with $B$ forming a structure of LP lobes. These oscillations occur due to the change with flux of the integer fluxoid number, $n(\Phi) = \lfloor \Phi/\Phi_0\rceil$, where $\Phi$ is the flux threading the shell, and $\Phi_0=h/2e$ is the superconducting flux quantum. We therefore label them as the zeroth (0L), first (1L) and second (2L) lobes. These features are maximum at fields where an integer number of flux quanta thread the shell (black dashed lines), and decay towards the edges of the lobes at half integer values of $\Phi_0$ (gray dashed lines). At lower current bias, we observe multiple Andreev reflection (MAR) features, specially in 0L, signaling a highly transparent junction. 

At low bias, the $dV/dI$ shows a temperature-broadened maximum, clearly visible in red, at $I \sim 100$~nA, which effectively provides an estimate of the critical current in the overdamped regime $I \sim I_c$ \cite{Tinkham:96, Ambegaokar:PRL69}.
As $B$ increases from $B=0$, $I_c$ is suppressed at $B \sim 70$~mT and re-emerges near $B \sim 90$~mT, and similarly around $B \sim 220$~mT and $B \sim 250$~mT. While $I_c$ in 0L decays from the lobe center at $B = 0$, it exhibits a distinct skewness in the other lobes. In 1L, it is skewed toward high fields (i.e., to the right within the lobe), and in 2L it is skewed toward low fields (i.e., to the left). Interestingly, in 0L and 1L, $I_c$ vanishes abruptly at the right gray dashed line, where the fluxoid number changes, and is then followed by a metallic region both in 1L and 2L. This is reminiscent of the destructive LP regime, where the LP effect is strong enough to drive the shell to the normal state around the lobe edges. However, the destructive regions should be finite at both sides of the gray dashed lines, which is not the case. These observations are more compatible with a shell in the non-destructive LP regime (with a finite gap at the lobe edges) and the presence of CdGM analogs below the gap, as we will argue later (see also Joule characterization in \SMref{SM:device}{S-I}).

To further characterize the sample, we measured $dV/dI$ as a function of $B$ and cryostat nominal temperature $T$ at zero current bias ($I = 0$), see Fig. \ref{fig:exp}(b). The higher-temperature features follow a modulation with $B$ that is symmetric with respect to the lobe centers, compatible with the LP effect of the shell. However, the critical temperature $T_c$ of the core, that is, the temperature above which the hybridized semiconductor becomes normal, mirrors the skewnesses of $I_c$, most prominently in 1L. The presence of skewness in $I_c$ and $T_c$ is reproducible across multiple devices (see \SMref{SM:repro}{S-II}).

To theoretically understand the previous experiments, we use the model schematically represented in Fig. \ref{fig:sketch}(b) and based on Ref.~\cite{Paya:PRB25}. The Ohmic junction between the SC shell and the SM core creates an electrostatic potential $U(r)$ for the nanowire electrons with a dome profile as a function the radial coordinate $r$ \cite{Vaitiekenas:S20}. For a chemical potential of the order of or below the dome maximum, the charge carriers are confined to a region near the SC/SM interface, occupying different CdGM analogs. Instead of discrete states as in Abrikosov vortices, these analogs are shell-induced Van Hove singularities in propagating one-dimensional core subbands \cite{San-Jose:PRB23}. Although the nanowire cross-section is hexagonal, an effective cylindrical model is a good approximation for describing the CdGM-analog physics with realistic Al/InAs parameters \cite{Paya:PRB24}. In this approximation, each CdGM analog has a good quantum number of the generalized angular momentum, allowing for efficient numerical calculations (see \ref{ap:model} and Ref.~\cite{Paya:PRB25} for details). Moreover, their wave function is typically centered at an average radius $R_{\rm{av}}$, which is smaller than the average shell radius $R_\text{LP}$. 

To perform microscopic numerical simulations, we need to know several geometric and intrinsic parameters of the shell and the core. Following the methodology of Ref.~\cite{Ibabe:NC23}, we independently determine the LP parameters for each full-shell lead on the left ($L$) and right ($R$) sides of the junction from measurements of $dI/dV$ as a function of voltage bias (see \SMref{SM:fullshell}{S-I.B}). This yields SC shell thicknesses $d^{L, R} = 8$~nm, radii $R^{L, R}_\text{LP} = 64$~nm (which translates into a nanowire radius $R_{\rm{NW}} = R_\text{LP} - d/2 = 60$~nm), pairing amplitudes $\Delta_0^{L, R} = 0.2$~meV, coherence lengths $\xi_d^L = 55$~nm and $\xi_d^R = 92$~nm and a magnetic field misalignment angle $\theta = 8^\circ$. These parameters place the full-shell wires close to, but not yet in, the destructive LP regime: the shell gap is suppressed near half-integer flux quanta, but it does not vanish. In this situation, the observed vanishing of $I_c$ at the lobe edges cannot be explained by a collapse of the shell gap alone. Instead, it must arise from a finite population of CdGM analogs crossing zero energy and thus closing their respective gaps in those $B$-field regions \cite{Paya:PRB25} \footnote{Notice that since both $R_{\rm{NW}}$ and $d$ are the same for both sides of the junction, there is no possible fluxoid mismatch phenomenology \cite{Paya:PRB25a}.}.


Concerning the SM nanowire, the electron effective mass is fixed to the standard value for Al/InAs nanowires, $m_0 = 0.023m_e$. Neglecting spin-orbit coupling, which is known to be weak in these devices \cite{Woods:PRB19} and non-essential for CdGM physics \cite{Paya:PRB24}, our microscopic model contains five free fitting parameters: the chemical potential $\mu$, the SC/SM coupling $\Gamma_{\rm{NS}}$, the junction transparency $T_N$, the average radius $R_\text{av}$ of the CdGM analog wave-functions and the effective $g$-factor of the core electrons. With all these parameters, we calculate the local density of states (LDOS) and the critical current using the Keldysh Green's function formalism \cite{A.A.Abrikosov:63, Martin-Rodero:PRL94, Perfetto:PRB09}, see \ref{ap:model} and Ref.~\cite{Paya:PRB25}.


\begin{figure}
\includegraphics[width=\columnwidth]{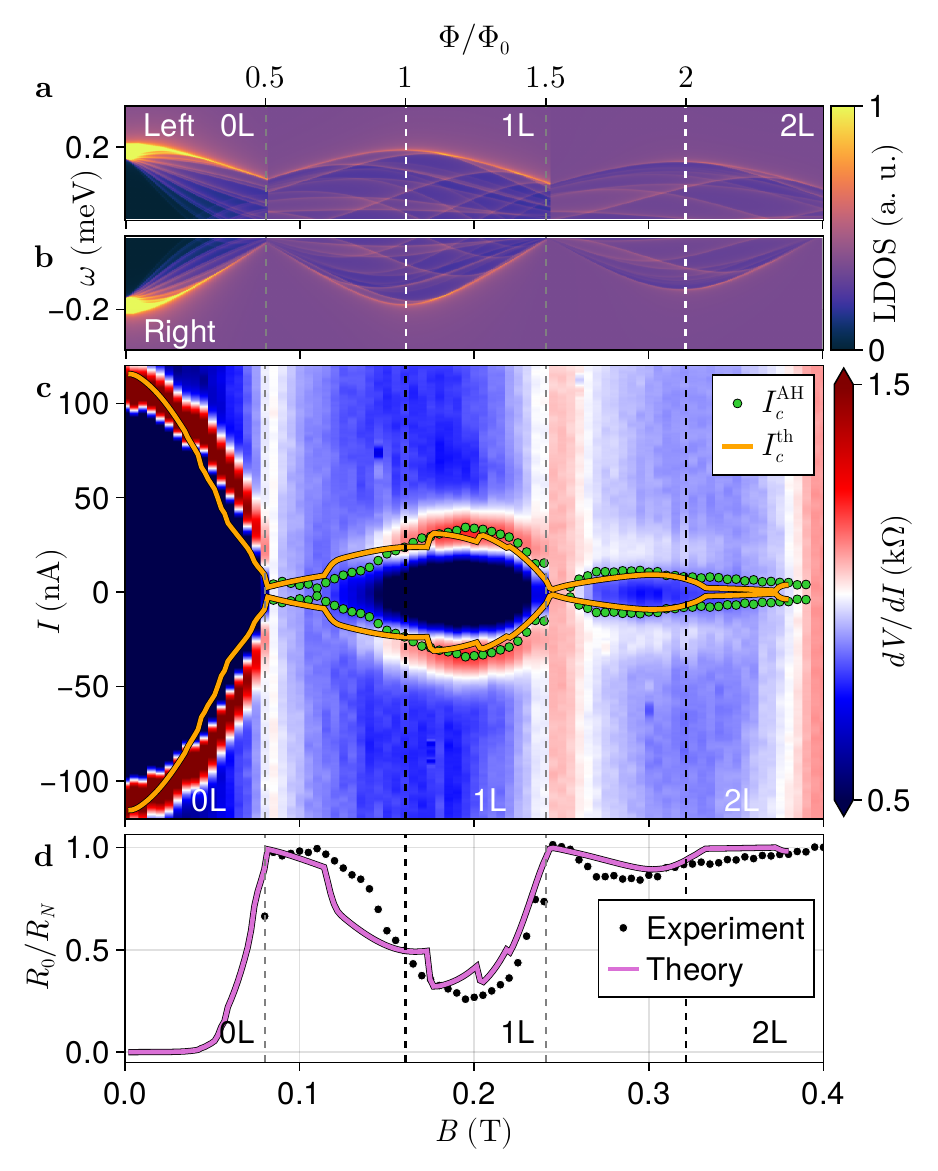}
\caption{\label{fig:theo}
\textbf{Critical current skewness caused by CdGM analogs.} Local density of states (LDOS) calculated at the end of a semi-infinite full-shell nanowire, corresponding to the left (a) and right (b) sides of the Josephson junction, as a function of magnetic field $B$ (or normalized flux $\Phi/\Phi_0$) and energy $\omega$. The shell gap edge is modulated into LP lobes ($0$L, $1$L, $2$L). Inside, CdGM analogs disperse with flux, exhibiting a distinct skewness toward high (low) magnetic fields within $1$L ($2$L). (c) Low-current zoom of the differential resistance map from Fig.~\ref{fig:exp}(a). The green dots indicate the critical current extracted from the experimental data using the Ambegaokar-Halperin formula, $I_c^\text{AH}$. The orange line represents the simulated critical current using the microscopic theoretical model, $I_c^\text{th}$. (d) Zero-current resistance $R_0$ normalized to normal resistance $R_N$ [extracted from (c)]. The black dots represent the experimental data, while the purple line is calculated using the Ambegaokar-Halperin formula with $I_c=I_c^\text{th}$.
Fitting parameters: $R_\text{av} = 47.5$~nm, $\mu = 20$~meV, $\Gamma_{\rm{NS}} = 2.5\Delta_0$, $g = 53$ and $T_N = 0.24$. The electronic temperature is found to be $T_e = 460$~mK.
}
\end{figure}


In Figs. \ref{fig:theo}(a, b) we show the LDOS at the ends of the $L$ and $R$ semi-infinite full-shell nanowires, respectively, as a function of magnetic field $B$ and energy $\omega$. They correspond to the LDOS on both sides of the normal junction with $T_N \rightarrow 0$ \footnote{Note that we only show the positive (negative) energy range for the $L$ ($R$) full-shell nanowire lead in Fig. \ref{fig:theo}(a) [Fig. \ref{fig:theo}(b)], since the LDOS is symmetric with respect to $\omega=0$. Note also that both LDOS profiles are very similar, since all hybrid wire parameters are the same except for the coherence lengths $\xi_d^L$ and $\xi_d^R$.}. The shell gap follows the expected LP modulation, with maxima (minima) at integer (half-integer) flux quanta through the shell. Within this gap, CdGM analogs appear, whose spectral density essentially covers the entire shell gap in 1L and 2L. Their Van Hove singularities, visible as a brighter color, disperse strongly with $B$ due mostly to the orbital effect of the magnetic field. Because the wave functions of these modes peak at $R_{\rm{av}}<R_{\rm{LP}}$, they effectively enclose a smaller area than the shell and therefore experience a different LP period. The threading-flux mismatch between the CdGM analogs and the shell effectively shifts the oscillation pattern of the subgap states relative to that of the parent gap. In addition, the Zeeman field breaks the spin degeneracy of the subgap states within each lobe. The combination of both effects skews the CdGM analogs with respect to the lobe centers. In 1L, the majority of CdGM-analog gaps are larger toward the right side of the lobe and cross zero energy steadily as one evolves toward the left side; the opposite happens in 2L. This does not happen in 0L, where the high symmetry point at $B = 0$ ensures that all CdGM states reach their maximum energy at that field. Because the zero-energy crossings occur at nearly identical fields on both sides of the junction, they open or close supercurrent channels in unison. As a consequence, $I_c$ inherits the skewness of the underlying subgap spectral density within each LP lobe. 


Figures \ref{fig:theo}(c, d) present a comparison between theory and experiments. In Fig. \ref{fig:theo}(c) we show a blow-up of Fig. \ref{fig:exp}(a) for small current bias, to highlight the behavior of the measured $I_c$. Although the resistance peak is clearly visible in red in 0L and 1L, it is less defined in 2L, making a direct fitting to theory challenging.
To help with this matter, in Fig. \ref{fig:theo}(d) we plot with black dots the zero-bias resistance $R_0$ (taken from Fig. \ref{fig:theo}(c) at $I=0$), normalized to the  normal resistance $R_N$, as a function of $B$. $R_N$ is defined as the resistance of the hybrid system above the critical current, and is extracted by averaging $dV/dI$ over the flat high-bias region above the red peak in Fig. \ref{fig:theo}(c). We do not provide $R_0/R_N$ for 0L because the strong MAR features in this B-field range prevent an accurate estimate of $R_N$ within this $B$-field range. However, this omission does not affect our analysis, as our primary focus is the skewness patterns of 1L and 2L.

In an ideal noiseless Josephson junction, one expects $R_0 = 0$. However, in the presence of thermal fluctuations, $R_0 /R_N$ is related to $I_c$ through the Ambegaokar-Halperin formula \cite{Ambegaokar:PRL69, Gross:16},
\begin{equation}\label{Eq:AH}
    \frac{R_0}{R_N} = \mathcal{I}_0^{-2}\left(\frac{\Phi_0}{2\pi k_B T_e} I_c\right),
\end{equation}
where $\mathcal{I}_0$ is the modified Bessel function of the first kind, $k_B$ the Boltzmann constant, and $T_e$ the electronic temperature. Since $\mathcal{I}_0^{-2}$ is a monotonic function, any skewness in $R_0 / R_N$ directly reflects the skewness of $I_c$. Notably, $R_0$ is influenced by both the subgap states and the shell gap edge, whereas $R_N$ is determined almost entirely by the latter. As a result, the ratio $R_0 / R_N$ effectively isolates the behavior of the CdGM analogs. 

Using Eq. \eqref{Eq:AH} and the experimental data in Fig. \ref{fig:theo}(d) (black dots), we extract the critical current $I_c^{\rm{AH}}$ and plot it as green dots in Fig. \ref{fig:theo}(c). To do so, we estimate $T_e$ by fitting $I_c^{\rm{AH}}$ to the clearly resolved $I_c$ peak in 1L\footnote{We take $T_e$ to be field-independent, as it is set by the thermal environment; indeed, the same value reproduces the 2L data without further adjustment.}. The solid orange curve represents the result of our microscopic simulation of the critical current, $I_c^\text{th}$, fitted to the green dots. The fitting parameters $\Gamma_{\rm{NS}}$ and $T_N$ are primarily determined by the $I_c$ maxima within each LP lobe, while $\mu$ and $R_\text{av}$ are governed by the $I_c$ skewness. 
These two parameters, however, leave distinct fingerprints on $I_c(B)$.
The chemical potential $\mu$ mainly sets which CdGM analogs are occupied, so varying it adds or removes subgap branches and changes $I_c$ in a predominantly step-like manner.
In contrast, $R_\text{av}$ sets the flux enclosed by the CdGM analogs and therefore smoothly tunes their LP mismatch with the shell, and hence the degree of skewness.
Since a single parameter set must reproduce the opposite skewness in 1L and 2L, as well as $R_0/R_N$, the two parameters cannot compensate for each other.
Additionally, the leftward skewness in 2L fixes the effective $g$-factor \footnote{Effective $g$-factors of subgap states ranging from $\sim 4$ to $\sim 50$ have been reported in Al/InAs nanowires~\cite{Vaitiekenas:PRL18}, stemming from the alignment of spin with the orbital angular momentum of higher subbands~\cite{Winkler:PRL17}. For reference, bulk InAs has $\vert{}g\vert{} \approx 15$. The enhanced value of $g \approx 53$ in our theoretical fits should be understood as an effective parameter tied to our tubular-core model, which constrains the radial profiles of all modes to peak at the same $R_\text{av}$. This constraint necessitates a larger $g$ to reproduce the skewness pattern of the 2L (for $g \lesssim 20$, the calculated 2L skewness points to the right). Conversely, the behavior of the critical current in the 0L and 1L is essentially independent of $g$, being determined primarily by $R_\text{av}$ and $\mu$.}.
For completeness, we substitute $I_c^\text{th}$ into Eq. \eqref{Eq:AH} and plot the resulting theoretical $R_0/R_N$ ratio as a solid purple curve in Fig. \ref{fig:theo}(d).



We observe remarkable agreement between the experimental magnetic field dependence of $I_c$ and the theoretical simulation $I_c^\text{th}$. We highlight that using a single set of fitting parameters, the behavior of both $I_c$ and $R_0/R_N$ is well captured across the entire $B$ range explored, including the distinct behaviors of 1L and 2L. This provides strong evidence that the experimentally observed skewness arises from CdGM analogs in the full-shell Al/InAs nanowires, in line with previous studies \cite{Deng:PRL25}. Indeed, repeating the calculation with a fully depleted semiconductor core, which leaves only the LP-modulated shell, yields a critical current that is symmetric within each LP lobe (see \SMref{SM:depleted}{S-III}). We note that rightward skewness in 1L is consistently observed across devices (see \SMref{SM:repro}{S-II}) and is always absent in 0L. In 2L, however, the skewness can be to either the right or the left depending on the parameters \cite{Paya:PRB25}, see \SMref{SM:repro}{S-II}. We also note that this overall behavior is consistent at different gate configurations, see \SMref{SM:repro}{S-II.C}.


In conclusion, we have investigated the anomalously skewed critical current $I_c$ consistently observed in overdamped Josephson junctions based on full-shell hybrid nanowires. This skewness appears only when the SC shell is threaded by a non-zero number of fluxoids, directly tying it to the LP effect of the shell. However, the skewness cannot be accounted for by a simple LP modulation of the shell gap. Through comparison with microscopic simulations, we show that the skewness is a consequence of subgap CdGM analogs residing in the SM core close to, but away from, the SC/SM interface. Because the wavefunctions of the CdGM analogs enclose a smaller area than the SC shell, they experience a different LP oscillation period, effectively skewing the dispersion of their induced gaps with flux within the LP lobes. This, in turn, results in a skewed $I_c$, which is the sum of the contributions of all SC shell and SM core supercurrent channels. The same skewness pattern is visible in the flux-modulated zero-bias resistance ($R_0/R_N$), which is linked to $I_c$ in overdamped junctions subject to thermal fluctuations. This work highlights the need to consider the contribution of the SM core, and the fact that the subgap wavefunction encloses a smaller area than the shell, when interpreting or calculating the critical current in realistic full-shell nanowire Josephson junctions.



\cpar{Acknowledgments} We acknowledge funding by the EU through the European Research Council (ERC) Starting Grant agreement 716559 (TOPOQDot), the FET-Open contract AndQC, the Danish National Research Foundation, Innovation Fund Denmark, the Carlsberg Foundation, and the Spanish AEI through Grants No.~PID2023-150224NB-I00, PRE2022-101362, PID2024-161156NB-I00, PID2024-161665NB-I00, and through the ``Mar\'{\i}a de Maeztu'' Program for Units of Excellence in R\&D (CEX2023-001316-M), and Severo Ochoa Centers of Excellence program (CEX2024-001445-S); funded by MICIU/AEI/10.13039/501100011033, ``ERDF A way of making Europe'' and ``ESF+''.

\cpar{Data availability} Data supporting the findings of this article and data used to analyze it are openly available in Ref.~\cite{Paya:26b}. Numerical calculations are based on \texttt{Quantica.jl} \cite{San-Jose:21}. Visualizations were made with the \texttt{Makie.jl} package \cite{Danisch:JOSS21}. 

\cpar{Author contributions} A.I. fabricated the device. A.I., M.G., and E.J.H.L. performed the measurements and analyzed the experimental data. C.P., P.S-J., R.A., and E.P. developed the theory. C.P. performed the theoretical calculations. T.K. and J.N. developed the nanowires. E.J.H.L. proposed and guided the experiment and E.P. proposed and guided the theory. All authors discussed the results. C.P., A.I., and E.P. wrote the manuscript with input from all authors.


\appendixpar{Sample fabrication} \label{ap:fabrication} The device studied in this work is based on InAs nanowires 
fully covered by an epitaxial Al shell with an estimated thickness $d \sim 8$~nm. The nanowires are deterministically transferred from the growth chip to Si/SiO$_2$ ($300$~nm) substrates using a micro-manipulator. E-beam lithography (EBL) is then used to define a window for wet etching a $\sim 200$~nm-long segment of the Al shell. A $30$~s descumming by oxygen plasma at $200$~W is performed before immersing the sample in the AZ326 MIF developer (containing $2.38$\% tetramethylammonium hydroxide, TMAH) for 60 s at room temperature. Electrical contacts and a side gate 
are subsequently fabricated by standard EBL techniques, followed by ion milling to remove the native oxide of the Al shell, and Ti ($2.5$~nm)/Al ($200$~nm) metallization by e-beam evaporation. The side gate allow us to tune the charge carrier density of the InAs nanowire. 

Notably, the main features discussed in this work have been reproduced in three additional full-shell nanowire Josephson devices, including  one measured in a four-terminal geometry (see \SMref{SM:fourterm}{S-II.B} for  experimental data from these devices).


\appendixpar{Measurements} \label{ap:measurements} The measurements presented in the main text were performed in a $^3$He cryogenic system with a base temperature of $250~\mathrm{mK}$. The DC wiring consisted of pi filters at room temperature, constantan twisted pairs down to the $^3$He pot, followed by low-temperature RC filters with a cut-off frequency  of $10~\mathrm{kHz}$. 

Transport measurements were performed in a two-terminal configuration using  standard lock-in techniques. The device was biased with a DC current and a  small AC excitation, while simultaneously measuring the voltage drop across  the device and the differential resistance. The AC current excitation was  kept in the range of $2.5-5~\mathrm{nA}$. A detailed description of the  measurement configuration and the procedure used to account for the finite  series resistance of the setup are provided in \SMref{SM:data}{S-I.A}. 

The parameters of  the full-shell nanowire leads used in the microscopic model were independently  extracted using Joule spectroscopy \cite{Ibabe:NC23}, as described in \SMref{SM:fullshell}{S-I.B}. Due to a slight experimental misalignment, the applied magnetic field $B$ was oriented at a small angle $\theta$ relative to the nanowire axis. To account for this, $\theta$ was treated as a free fitting parameter in the Joule spectroscopy characterization.

Additional measurements performed on other devices, including measurements  in a dilution refrigerator system and a device measured in a four-terminal  configuration, are presented in \SMref{SM:twoterm}{S-II.B}. 

\raggedbottom
\appendixpar{Model} \label{ap:model}
The observables presented in this work are calculated within the Keldysh Green's functions formalism \cite{A.A.Abrikosov:63}. We model the system as two semi-infinite full-shell nanowires described by a tight-binding Bogoliubov-de Gennes (BdG) Hamiltonian. These nanowires are coupled through $\mathcal{V}$, defined as a fraction of the intra-wire hopping. Letting $g^r$ denote the retarded Green's function of the uncoupled nanowire, the tunneling  self-energy is $\Sigma^r = 
\mathcal{V}^\dagger g^r \mathcal{V}$. The dressed retarded Green's function within the first unit cell of the wire is the given by the Dyson equation
\begin{equation}
    G^r = \left[\left(g^r\right)^{-1} - \Sigma^r\right]^{-1}.
\end{equation}
The LDOS is therefore
\begin{equation}
    \rho = -\frac{1}{\pi} \text{Im} \left[\text{Tr} \left(G^r\right)\right].
\end{equation}
Furthermore,  the equilibrium Josephson current as a function of the phase difference $\phi$ between the nanowires reads
\begin{equation}
    J_S(\phi) = \frac{2e}{h} \text{Re} \int{d\omega f(\omega) \text{Tr}\left[\left(\Sigma^r G^r - G^r\Sigma^r\right)\tau_z\right]},
\end{equation}
where $f(\omega)$ is the Fermi-Dirac distribution and the trace is taken over spin and electron-hole degrees of freedom. Here, $\tau_z$ is the 3rd Pauli matrix in electron-hole space. Finally, the critical current is defined as $I_c = \text{max}\left[J_S(\phi)\right]$.

We \pagebreak refer to the Appendix in Ref.~\cite{Paya:PRB25} for a brief discussion of the BdG Hamiltonian.
We use the tubular-core approximation, in which the radial profile of the eigenvectors is assumed to be $\Psi(r) = \delta(r - R_\text{av})$ and $\mu - \langle U(r) \rangle \to \mu$. This approximation replaces the smooth, dome-shaped electrostatic potential created by the Ohmic SC/SM contact with a single effective radius at which the CdGM-analog wavefunctions are concentrated. As shown in Refs.~\mbox{\cite{San-Jose:PRB23, Paya:PRB24, Paya:PRB25}}, it reproduces the subgap phenomenology of the full radial model as long as $R_\text{NW}/2 < R_\text{av} < R_\text{NW}$, a condition satisfied by our fitted value $R_\text{av} = 47.5$~nm ($R_\text{NW} = 60$~nm). Including the full radial potential would introduce only quantitative corrections to the fitted parameters, without altering the origin of the skewness. Refs.~\mbox{\cite{San-Jose:PRB23, Paya:PRB24}} provide an in-depth discussion of the Hamiltonian.

\bibliography{Josephson, supp}



\ifSMincluded
  \onecolumngrid
  \clearpage
  \afterpage{%
    \setcounter{page}{1}%
    \renewcommand{\thepage}{S\arabic{page}}%
  }%
  \setcounter{figure}{0}
  \setcounter{table}{0}
  \setcounter{equation}{0}
  \setcounter{section}{0}
  \renewcommand{\thetable}{S\arabic{table}}
  \renewcommand{\thefigure}{S\arabic{figure}}
  \setcounter{secnumdepth}{3}
  \renewcommand{\thesection}{S\arabic{section}}
  \renewcommand{\thesubsection}{\Alph{subsection}}
  \begin{center}\textbf{\large Supplemental Material}\end{center}
  \vspace{0.5em}
  \FloatBarrier
  \input{SM_content.tex}

\fi

\end{document}

%% file: SM_content.tex
\section{Device parameters}
\label{SM:device}
\subsection{Data processing}
\label{SM:data}

\begin{figure}[h]
\centering
\includegraphics{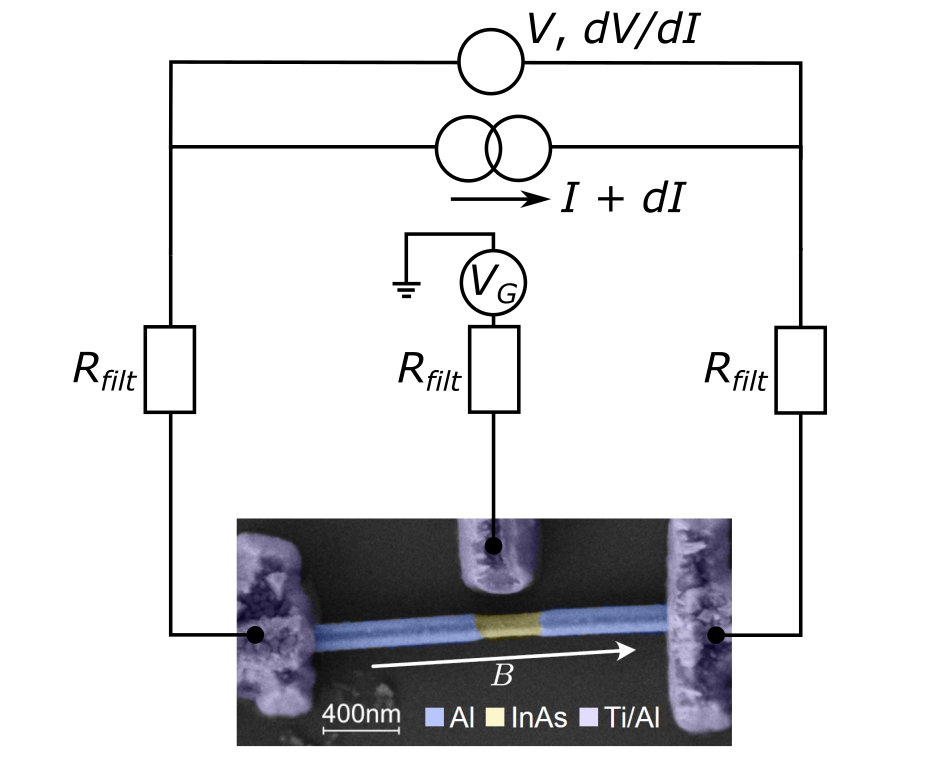}
\caption{
\textbf{Schematic of the measurement configuration.}
Two-terminal current-biased transport measurements are performed by applying 
a DC current bias with a small AC excitation, $I+dI$, while simultaneously 
measuring the voltage drop $V$ and differential resistance $dV/dI$ using 
standard lock-in techniques. The measured voltage and differential resistance include the contribution 
from the series resistance of the measurement lines, dominated by the 
cryogenic low-pass filters $R_{\rm filt}$. The local gate 
voltage $V_G$ controls the junction transparency.
}
\label{fig:circuit}
\end{figure}

The transport measurements presented in the main text were performed in a 
two-terminal configuration (Fig.~\ref{fig:circuit}). As a consequence, the measured voltage drop 
contains contributions from both the device and a finite series resistance 
$R_s$ arising from the measurement setup. The dominant contribution to $R_s$ originates from the low-pass filters 
installed in the cryogenic wiring, $R_{\rm filt}$. The total series resistance 
can be written as $R_s=2R_{\rm filt}+R_{\rm other}$, where $R_{\rm other}$ 
accounts for additional contributions from the remaining wiring and contact 
resistances.

To obtain the intrinsic voltage drop $V$  across the Josephson junction, we 
subtract this contribution from the measured voltage $V_{\mathrm{meas}}$ according to
\[
V=V_{\mathrm{meas}}-IR_s .
\]
The value of $R_s$ was determined experimentally by using the multiple 
Andreev reflection (MAR) resonances of the junction as an internal voltage 
calibration. MAR processes occur at well-defined energies
\[
eV=\frac{2\Delta_S}{m},
\]
where $\Delta_S$ is the superconductor shell gap at $B=0$ and $m$ is a natural number. Thus, the  MAR resonance positions in voltage should remain independent of gate voltage $V_G$, apart from changes in intensity associated with variations of the junction transparency. We therefore choose $R_s$ such that the MAR resonances become horizontal in voltage when plotting the differential conductance $dI/dV$ as a 
function of the corrected voltage bias $V$.

\begin{figure}[h]
\centering
\includegraphics[width=0.55\linewidth]{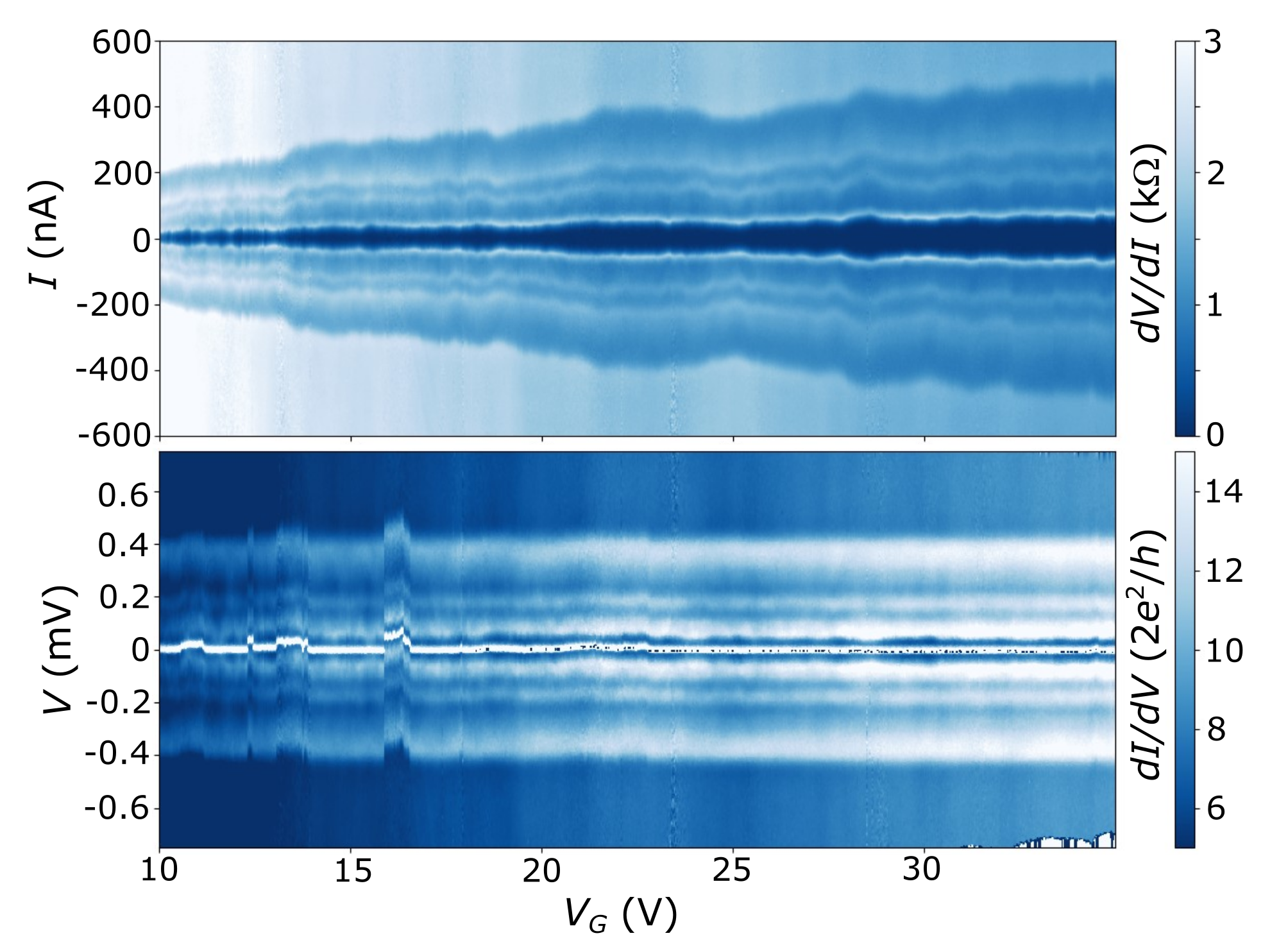}
\caption{
\textbf{Gate-voltage dependence of the main device.}
(a) Differential resistance $dV/dI$ measured as a function of applied current 
bias $I$ and gate voltage $V_G$ at zero magnetic field. A finite series 
resistance $R_s=5490~\Omega$, arising mainly from the cryogenic low-pass 
filters and wiring in the two-terminal measurement setup, has been subtracted.
(b) Differential conductance $dI/dV$ plotted as a function of the corrected 
voltage drop across the junction $V$ after subtracting $R_s$. With this correction, multiple Andreev reflection (MAR) 
features remain approximately constant in $V$ as a function of $V_G$, providing an internal calibration of the voltage drop across the junction. The corrected data also yield an approximately vanishing zero-bias resistance in the superconducting state.
}
\label{fig:gate}
\end{figure}

Using this procedure, we obtain $R_s=5490~\Omega$ for the main device. As 
shown in Fig.~\ref{fig:gate}, after this correction the MAR resonances appear 
at approximately constant voltage over the whole $V_G$ range. Consistently, the corrected zero-bias resistance in the superconducting state 
at $B=0$ is reduced to approximately zero, as expected for a Josephson junction. The same value of $R_s$ is used for the correction of all 
experimental data presented in the main text.

\pagebreak
\subsection{Extracting full-shell parameters with Joule spectroscopy}
\label{SM:fullshell}



\begin{figure}[h]
\centering
\includegraphics[width=0.65\linewidth]{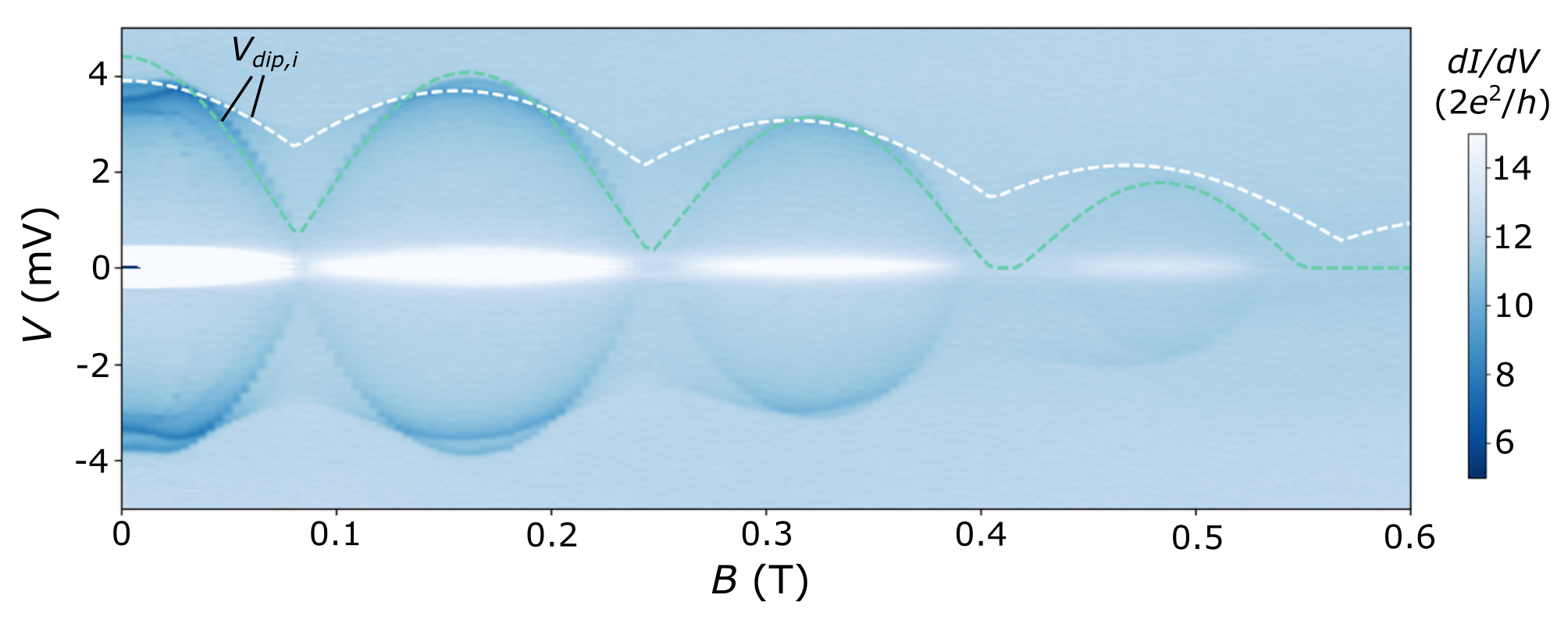}
\caption{
\textbf{Joule spectroscopy characterization of the main device.}
Differential conductance $dI/dV$ as a function of voltage drop $V$ and magnetic field $B$. At high bias, Joule heating drives the superconducting shells into the normal state, producing conductance dips at $V_{\rm dip,i}$ associated with the left and right 
superconducting leads. The dip positions 
follow the LP modulation of the superconducting critical temperature and are fitted as in Ref.~\cite{Ibabe:NC23} (white and green dashed lines for $L$ and $R$ leads, respectively). The resulting parameters are 
used as inputs for the microscopic model presented in the main text.
}
\label{fig:joule}
\end{figure}

The parameters of the Al full-shell leads used in the microscopic model presented in the main text were extracted using Joule spectroscopy. This method detects the superconductor-to-normal metal transition of the superconducting leads by Joule heating through the suppression of the excess current, which manifests as dips in the differential conductance $dI/dV$, as shown in Fig.~\ref{fig:joule}. 

As described in detail in Ref.~\cite{Ibabe:NC23}, within the Joule 
spectroscopy framework, the position of these conductance dips, 
$V_{\rm dip,i}$, is proportional to the superconducting critical temperature 
of each shell, $T_{c,i}$, where $i=L,R$ denotes the left and right leads, 
respectively. The magnetic-field dependence of $V_{\rm dip,i}(B)$ therefore 
provides a direct probe of the Little-Parks (LP) modulation of the superconducting 
state. By fitting $V_{\rm dip,i}(B)$ using the theory in Ref.~\cite{Ibabe:NC23}, we extract the effective shell radius $R_{\rm LP}^{i}$, 
the superconducting coherence length $\xi_{d}^{i}$, and the angle $\theta^i$ between 
the external magnetic field and the nanowire axis, as shown in Table~\ref{tab:param}.

\begin{table}[h]
\centering
\caption{
\textbf{Full-shell parameters extracted from Joule spectroscopy.}
Parameters obtained by fitting $V_{\rm dip,i}(B)$. The white and green dashed lines in Fig.~\ref{fig:joule} correspond to the left $L$ and right $R$ shell leads, respectively.
}
\label{tab:joule}
\begin{tabular}{c c c c c}
\hline
Lead & $d$ (nm) & $R_{\rm LP}$ (nm) & $\xi_d$ (nm) & $\theta$ ($^\circ$) \\
\hline
$L$ & 8 & 64 & 55 & 8 \\
$R$ & 8 & 63.5 & 92 & 5 \\
\hline
\end{tabular}
\label{tab:param}
\end{table}

\pagebreak
\section{Reproducibility of the critical current skewness}
\label{SM:repro}
To investigate the reproducibility of the observed phenomenology, we measured 
several full-shell nanowire Josephson junctions under an axial magnetic field. 
A total of four devices showed qualitatively similar behavior, characterized 
by an asymmetric evolution of the critical current $I_c$ within the nonzero LP lobes. Here, we present additional measurements from two representative 
devices, demonstrating that the $I_c$ skewness reported in the 
main text is not specific to a single junction.

\subsection{Additional two-terminal devices}
\label{SM:twoterm}

\begin{figure}[h]
\centering
\includegraphics[width=0.9\linewidth]
{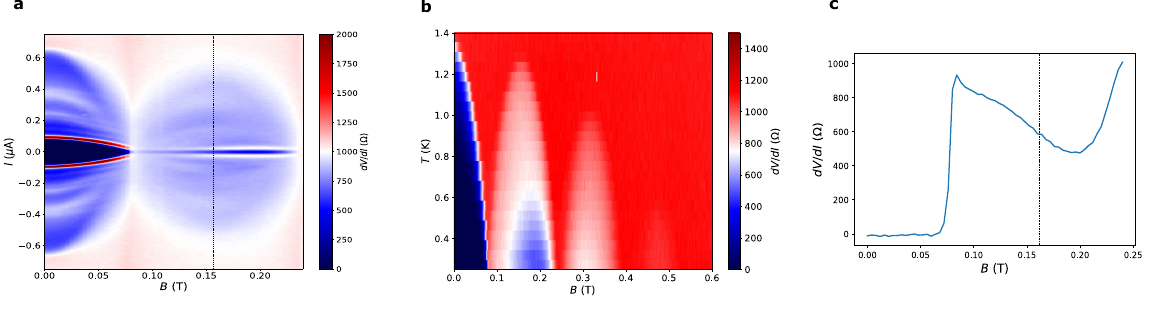}
\caption{
\textbf{Additional data for Device B.}
(a) Differential resistance $dV/dI$ as a function of current bias, $I$, and applied magnetic field, $B$. (b) Zero-bias $dV/dI$ measurement as a function of $B$ and of the bath temperature, $T$. (c) Zero-bias differential resistance, $R_0$, as a function of $B$ taken from (a). The dashed line is a guide to the eye indicating the center of the first LP lobe.}
\label{fig:SIB}
\end{figure}

Figure~\ref{fig:SIB} shows data from an additional two-terminal device 
(Device B) fabricated on the same chip and measured during the same cool-down 
as the device discussed in the main text. The device shows a similar critical 
current skewness, with the maximum $I_c$ shifted away from the center of the first LP lobe. Following the procedure described 
above, we subtract a series resistance $R_s=5490~\Omega$.

\begin{figure}[h]
\centering
\includegraphics[width=0.85\linewidth]
{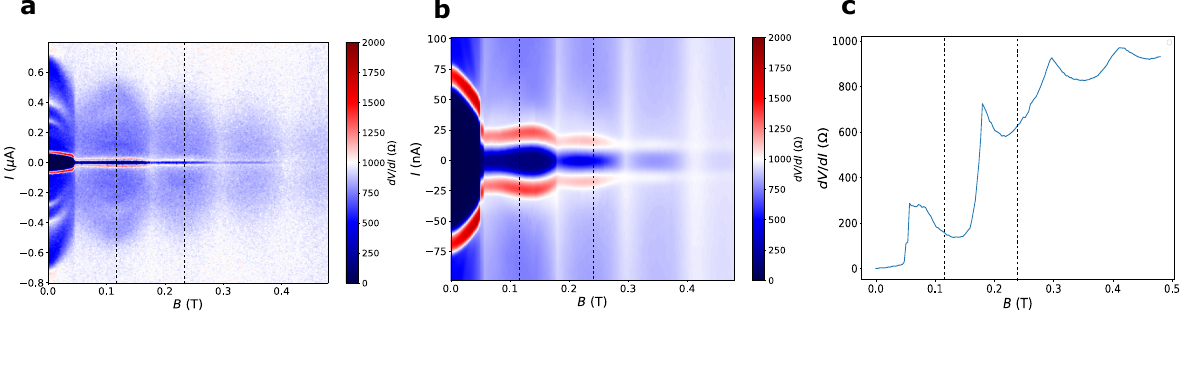}
\caption{
\textbf{Additional data for Device C.}
(a) Differential resistance $dV/dI$ measured as a function of current bias $I$ and external magnetic field  $B$, revealing MAR features. 
(b) Same as (a) but focusing on the critical current features. (c) Zero-bias differential resistance, $R_0$, as a function of $B$ taken from (b). Dashed lines correspond to the center of the first and second LP lobes.
}
\label{fig:SIC}
\end{figure}

Data from another two-terminal device (Device C) are shown in Fig.~\ref{fig:SIC}. In this case, the obtained shell effective radius is larger than that of the device in the main text, $R_{\rm LP}\sim$ 77 nm, resulting in a reduced LP periodicity ($\sim 110$~mT). A series resistance $R_s=5350~\Omega$ was subtracted following the same procedure described above.

\subsection{Four-terminal device}
\label{SM:fourterm}
Device D corresponds to a four-terminal full-shell nanowire Josephson junction, 
where normal Cr/Au electrodes were used instead of superconducting Ti/Al 
contacts. Due to the four-terminal measurement configuration, the data were 
not processed to subtract a series resistance. Consistently, the differential resistance of the supercurrent branch at $B=0$ is approximately zero without any correction. Despite the different device geometry, Device D exhibits the same qualitative behavior discussed in the main text, featuring a rightward shift of the $I_c$ maximum relative to the maximum of the LP gap modulation in the first LP lobe.

We note that a larger misalignment angle between the applied magnetic field and the nanowire axis in this device results in a stronger orbital suppression of superconductivity with increasing $B$. As a consequence, the accessible field 
range is mainly limited to the zeroth and first LP lobes, preventing an analysis of the second LP lobe, as performed for the main device.

\begin{figure}[h]
\centering
\includegraphics
{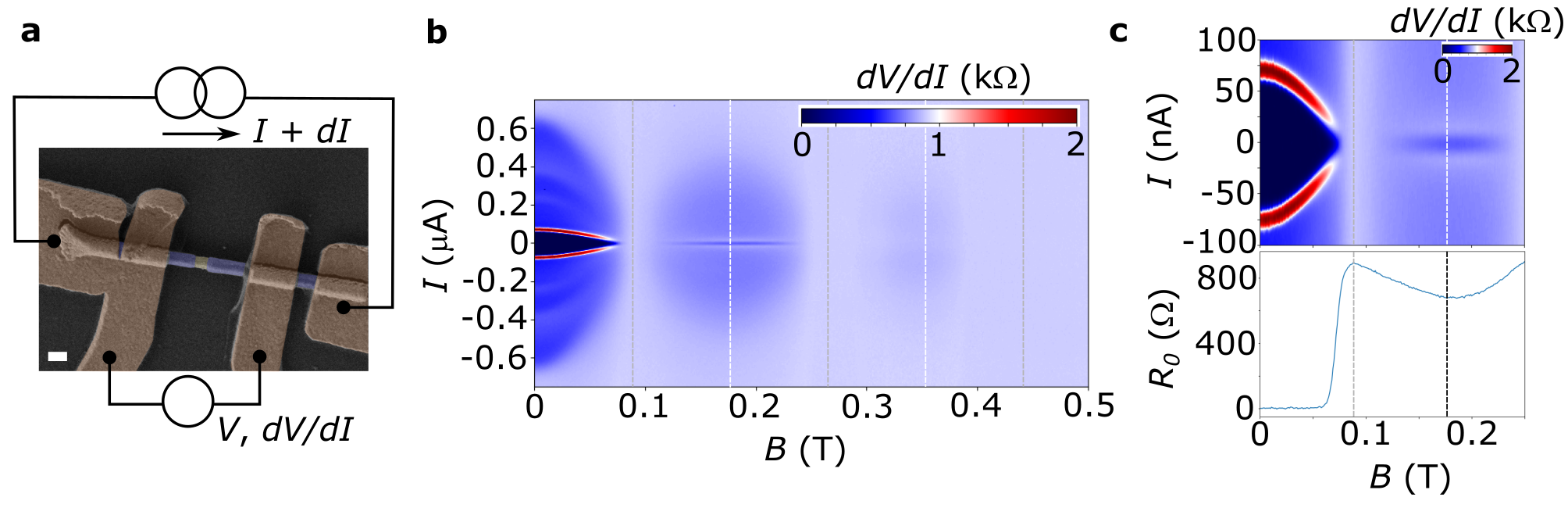}
\caption{
\textbf{Critical current skewness in a four-terminal device.}
(a) False-colored scanning electron micrograph of Device D, consisting of a 
full-shell InAs-Al nanowire Josephson junction contacted by normal Cr/Au 
electrodes in a four-terminal configuration. 
(b) Differential resistance, $dV/dI$, measured at $T=250~\mathrm{mK}$ using 
an AC excitation $dI=10~\mathrm{nA}$, as a function of current bias $I$ 
and magnetic field $B$. 
(c) Zoom-in of the $dV/dI(I,B)$ measurement in the zeroth and first LP lobes 
(top), and zero-bias differential resistance, $R_0$, as a function of $B$ (bottom), highlighting the critical current skewness in Device D.}
\label{fig:devD}
\end{figure}

\subsection{Device A at different gate voltages}
\label{SM:gates}
\begin{figure}[h]
\centering
\includegraphics[width=\textwidth]{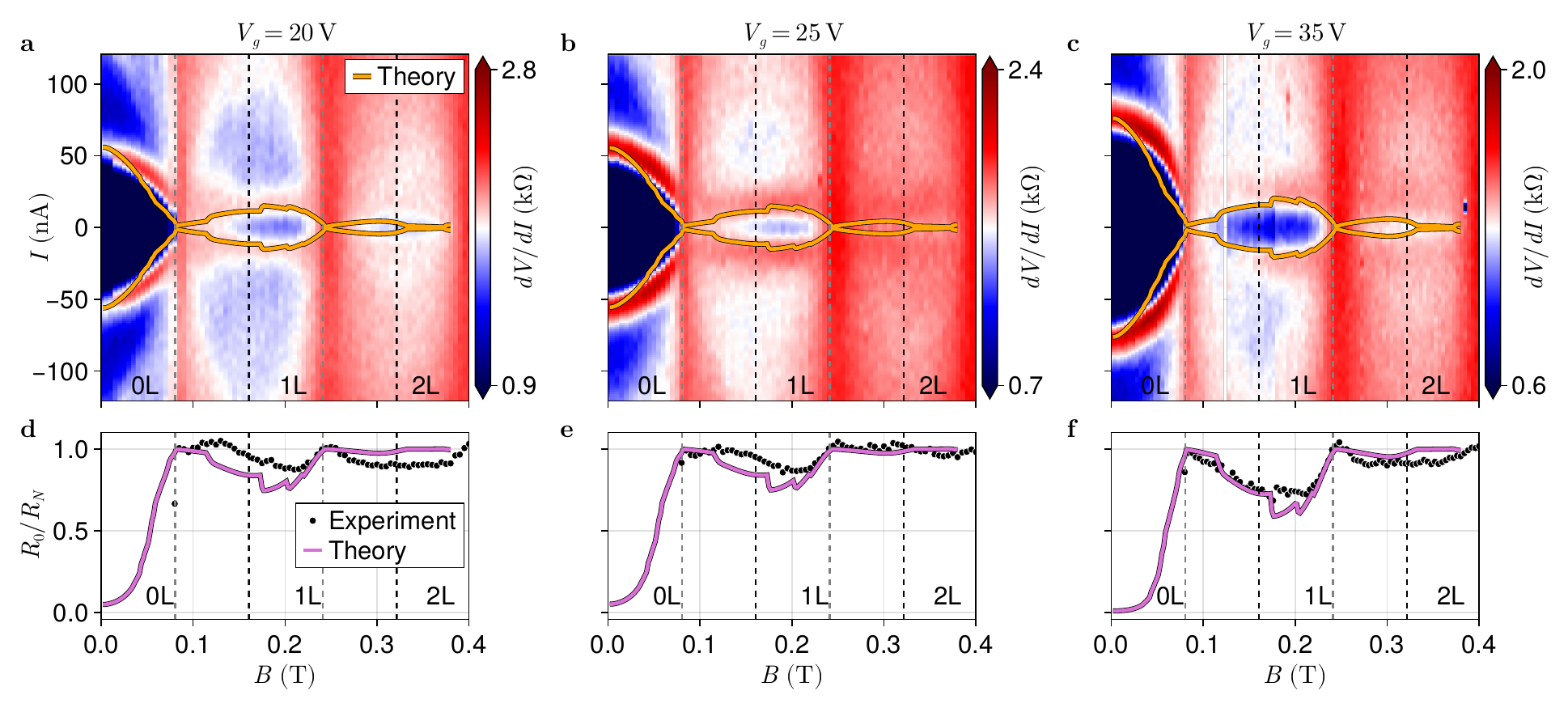}
\caption{\textbf{Critical current at different gate voltages.} $dV/dI$ measurements as a function of $I$ and $B$ for $V_G = 20 V (a)$, $V_G = 25 V (b)$ and $V_G = 35 V (c)$. The theoretical curves (yellow) corresponds to the same curves as in the main text, but they are rescaled to match the critical current at zero field. (d-e) Extracted zero-bias $R_0/R_N$ and compared to the $I_c^\text{th}$.}
\label{fig:gates}
\end{figure}

The $dV/dI$ measurements as a function of magnetic field presented in the main text correspond to the open regime where $V_G = 80V$. Figure~\ref{fig:gates} shows $dV/dI$ measurements obtained at lower gate voltages, namely $V_G = 20$ V (a), $V_G = 25$ V (b) and $V_G = 35$ V (c). They correspond to average normal resistances $\langle R_N \rangle \sim 1.6$k$\Omega$, $\langle R_N \rangle \sim 1.4$k$\Omega$ and $\langle R_N \rangle \sim 1.2$k$\Omega$, respectively, to be compared with the $\langle R_N \rangle = 0.7$k$\Omega$ of Fig.~2 of the main text. For their corresponding theoretical critical current curves in yellow ($I_c^\text{th}$), we have rescaled the critical current obtained at $V_G = 80$ V by a factor of $0.4$, $0.4$ and $0.5$, respectively, while keeping the others extracted parameters the same, which effectively results in a lower $T_N$ as the gate voltage is reduced. Following the same logic as in the main text, we compare the zero-bias $R_0/R_N$ with the theoretical curve (purple) for each of the measurements (Figs ~\ref{fig:gates}.d-f), rescaled to account for the reduced $T_N$.

\section{Depleted vs. filled hybrid nanowire}
\label{SM:depleted}

\begin{figure}[h]
\centering
\includegraphics[width=0.65\linewidth]{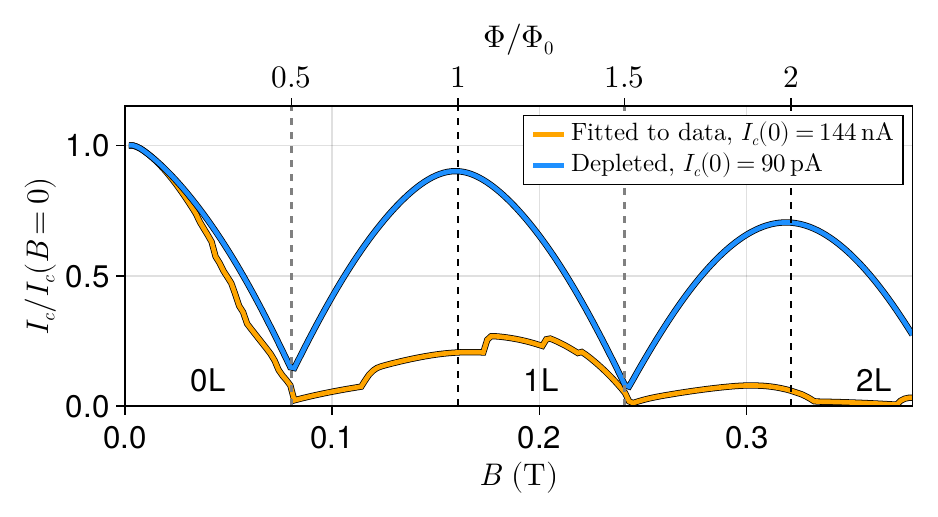}
\caption{\textbf{Fitted vs. depleted critical current.}
Calculated critical current $I_c$, normalized to its zero-field value, as a function of magnetic field $B$ (bottom axis) and normalized flux $\Phi/\Phi_0$ (top axis).
The orange curve is the fitted result of Fig.~3 of the main text, and the blue curve is the same calculation with the semiconductor core fully depleted, so that no CdGM analogs are present.
All other parameters are identical.
Vertical dashed lines mark integer (black) and half-integer (gray) flux quanta.}
\label{fig:depleted}
\end{figure}

Within the numerical model, the LP contribution to the critical current can be isolated by fully depleting the semiconductor core.
In this case, the supercurrent can flow only through the above-gap continuum of states, which strongly reduces its magnitude.
In Fig.~\ref{fig:depleted}, we compare the fitted critical current of Fig.~3 of the main text (orange) with that of the same system with a depleted core (blue).
Since the zero-field critical current of the depleted system is about three orders of magnitude smaller than that of the fitted one ($90$~pA vs.\ $144$~nA), we normalize each curve to its own value at $B = 0$ to allow a direct comparison.
The blue curve is symmetric about the center of each LP lobe, showing that the LP modulation of the shell gap alone does not produce any skewness.
This confirms that the skewness of the orange curve, and hence of the measured $I_c$, arises solely from the CdGM analogs that populate the superconducting gap of both nanowire leads.